\documentclass[twocolumn,floatfix, showpacs]{revtex4}

\usepackage[final]{epsfig}
\usepackage{amsmath}
\usepackage{parskip}
 
\begin{document}
 
\title{The Phenomenology of Elastic Energy Loss}
 
\author{Thorsten Renk}
\email{trenk@phys.jyu.fi}
\affiliation{Department of Physics, P.O. Box 35 FI-40014 University of Jyv\"askyl\"a, Finland}
\affiliation{Helsinki Institute of Physics, P.O. Box 64 FI-00014, University of Helsinki, Finland}
 

\begin{abstract}
The unexpectedly strong suppression of high $p_T$ heavy-quarks in heavy-ion collisions has given rise to the idea that partons propagating through a medium in addition to energy loss by induced radiation also undergo substantial energy loss due to elastic collisions. However, the precise magnitude of this elastic energy loss component is highly controversial. While it is for a parton inside a medium surprisingly difficult to define the difference between elastic and radiative processes rigorously, the main phenomenological difference is in the dependence of energy loss on in-medium pathlength: in a constant medium radiative energy loss is expected to grow quadratically with pathlength, elastic energy loss linearly. In this paper, we investigate a class of energy loss models with such a linear pathlength dependence and demonstrate that they are incompatible with measured data on hard hadronic back-to-back correlations where a substantial variation of pathlength is probed. This indicates that any elastic energy loss component has to be small.
\end{abstract}
 
\maketitle

\section{Introduction}

The energy loss of hard partons propagating through the soft medium created in heavy-ion collisions has long been regarded as a promising tool to gain information on the medium density evolution \cite{Jet1,Jet2,Jet3,Jet4,Jet5,Jet6}. Radiative energy loss, i.e. the idea that medium-induced radiation predominantly carries away energy from a hard parent parton has been rather successful in describing not only the hadronic nuclear suppression factor for central collisions but also the effects of changing medium geometry \cite{Dainese}. Calculations within dynamical evolution models in various formalisms \cite{Dyn01,Dyn02,Dyn03} have improved on this result and show also agreement with measured hard back-to-back correlations \cite{Correlations1,Correlations2,Correlations3} and the measured suppression of protons \cite{ppbar}.

Neverteless, there are indications that a radiative energy loss picture fails to describe the suppression of heavy quarks as seen in the single-electron spectra measurements \cite{HQPuzzle} and energy loss due to elastic collisions with the medium \cite{Mustafa,Mustafa2,DuttMazumder,Djordjevic,Wicks} has been suggested as a possible solution to this problem. Such calculations indicate a large component of elastic energy loss also for light quarks and gluons.

In vacuum, the distinction between an elastic and a radiative process is straightforward --- if the number of asymptotic out-states is larger than the number of asymptotic in-states, the process is radiative. However, for a parton propagating through a medium, no asymptotic out-states can be defined. In particular, radiated quanta do not need to be on-shell as long as they re-interact with the medium within a sufficiently short amount of timescale as set by the uncertainty principle. It becomes thus to some degree a matter of convention if a particular process is seen as the (elastic) exchange of a virtual parton between hard parton and medium or as the (inelastic) radiation of a virtual parton from the parent where the radiated parton is subsequently absorbed by the medium.

However, while there is no sharp conceptual distinction, there are two crucially different regimes: If a parton radiated from the parent is highly spacelike, the formation time for this process is very short. On the other hand, the decoherence time for near on-shell radiation is very long, giving rise to interference effects and LPM suppression. Phenomenologically, the first regime leads (in a medium with constant density) to a linear dependence of energy loss on pathlength whereas in the latter regime a quadratic dependence appears.

It is the purpose of this paper to study observable consequences of this difference in pathlength dependence. To this end, we first construct a class of models in which the energy loss has parametrically a linear pathlength dependence. Based on the requirement that the nuclear suppression factor $R_{AA}$ for central collisions should be described, we point out limits for the parameter space of this model class. Finally, we demonstrate with three different scenarios the consequences of linear pathlength dependence for observables which explicitly probe pathlength dependence such as $R_{AA}$ vs. reaction plane or hard back-to-back dihadron correlations and discuss limits for the relative magnitude of an elastic energy loss component.

\section{Modelling Energy Loss}

Key quantity for the calculation of energy loss in a dynamically evolving medium is the probability distribution $P(\Delta E; E)_{path}$ for a parton with initial energy $E$ to lose the energy $\Delta E$ for any given path through the medium (in general, this includes a discrete part accounting for the possibility that the parton escapes without energy loss). Empirically, one finds that a strong dependence of $P(\Delta E; E)_{path}$ on the initial energy $E$ does not seem to be favoured by the data \cite{gamma-hadron}, thus we will in the following approximate $P(\Delta E; E)_{path} \approx P(\Delta E)_{path}$.

From the energy loss distribution given a single path, we can define the overlap-geometry averaged energy loss probability distribution $\langle P(\Delta E)\rangle_{T_{AA}}$ as
\begin{equation}
\label{E-P_TAA}
\langle P(\Delta E)\rangle_{T_{AA}} \negthickspace = \negthickspace \frac{1}{2\pi} \int_0^{2\pi}  
\negthickspace \negthickspace \negthickspace d\phi 
\int_{-\infty}^{\infty} \negthickspace \negthickspace \negthickspace \negthickspace dx_0 
\int_{-\infty}^{\infty} \negthickspace \negthickspace \negthickspace \negthickspace dy_0 P(x_0,y_0)  
P(\Delta E)_{path}.
\end{equation}

Here, $\phi$ is the angle of the outgoing parton with respect to the reaction plane and
 $P(x_0, y_0)$ is the probability density for finding a hard vertex at the 
transverse position ${\bf r_0} = (x_0,y_0)$ given the collision impact 
parameter ${\bf b}$. This quantity is given by the product of the nuclear profile functions as
\begin{equation}
\label{E-Profile}
P(x_0,y_0) = \frac{T_{A}({\bf r_0 + b/2}) T_A(\bf r_0 - b/2)}{T_{AA}({\bf b})},
\end{equation}
where the thickness function is given in terms of Woods-Saxon the nuclear density
$\rho_{A}({\bf r},z)$ as $T_{A}({\bf r})=\int dz \rho_{(A}({\bf r},z)$. 

If we are interested in more differential observables, the relevant averaging procedure is changed, for example $R_{AA}$ as a function of the reaction plane is obtained from

\begin{equation}
\label{E-P_phi}
\langle P(\Delta E)\rangle_\phi \negthickspace = 
\int_{-\infty}^{\infty} \negthickspace \negthickspace \negthickspace \negthickspace dx_0 
\int_{-\infty}^{\infty} \negthickspace \negthickspace \negthickspace \negthickspace dy_0 P(x_0,y_0)  
P(\Delta E)_{path}
\end{equation}

or the back-to-back per-trigger-yield can be computed using $\langle P(\Delta E,E)\rangle_{Tr}$ which is found by replacing $P(x_0,y_0)$ in Eq.~(\ref{E-P_TAA}) by $P_{Tr}(x_0,y_0, {\bf p})$, the conditional probability density to find a vertex in the transverse plane from which a near side hadron in the relevant trigger range was produced. The latter quantity is best calculated in a Monte-Carlo (MC) framework as it also has explicit momentum dependence, for details see \cite{Correlations2}.  

We calculate the momentum spectrum of hard partons in leading order perturbative QCD (LO pQCD) (explicit expressions are given in \cite{Correlations2} and references therein). For reasonably hard momenta (in practice 6 GeV or more) it can be assumed that hadronization takes place outside the medium. Then the medium-modified observables arise from a convolution of the pQCD parton spectrum with the suitably averaged energy loss probability density and a hadronization function. For example, the medium-modified perturbative production of hadrons at angle $\phi$ can be computed from $\langle P(\Delta E)\rangle_\phi$ and the partonic cross section $\frac{d\sigma_{vac}^{AA \rightarrow f +X}}{d \phi}$ using
\begin{equation}
\frac{d\sigma_{med}^{AA\rightarrow h+X}}{d\phi} \negthickspace = \sum_f \frac{d\sigma_{vac}^{AA \rightarrow f +X}}{d\phi} \otimes \langle P(\Delta E)\rangle_\phi \otimes
D_{f \rightarrow h}^{vac}(z, \mu_F^2)
\end{equation} 
with $D_{f \rightarrow h}^{vac}(z, \mu_F^2)$ the fragmentation function for parton $f$ with momentum fraction $z$ at scale $\mu_F^2$ \cite{KKP,AKK}. From this we compute the nuclear modification function $R_{AA}$ vs. reaction plane as
\begin{equation}
R_{AA}(p_T,y,\phi) = \frac{dN^h_{AA}/dP_Tdy d\phi}{T_{AA}({\bf b}) d\sigma^{pp}/dP_Tdy d\phi}.
\end{equation}

Thus, all the information about the medium and the energy loss model is contained in $P(\Delta E; E)_{path}$. For radiative energy loss, we have in past works obtained this quantity by evaluating the two line integrals

\begin{equation}
\label{E-omega}
\omega_c({\bf r_0}, \phi) = \int_0^\infty d \xi \xi \hat{q}(\xi)
\end{equation}
and
\begin{equation}
\label{E-qL}
\langle\hat{q}L\rangle ({\bf r_0}, \phi) = \int_0^\infty d \xi \hat{q}(\xi)
\end{equation} 

along the path of the parton through the medium and using the results of \cite{QuenchingWeights} to convert the result to a probability distribution where we have assumed \cite{Flow1,Flow2}

\begin{equation}
\label{E-qhat}
\hat{q}(\xi) = K \cdot 2 \cdot \epsilon^{3/4}(\xi) (\cosh \rho(\xi) - \sinh \rho(\xi) \cos\alpha).
\end{equation}

Both the local energy density $\epsilon(\xi)$ and the transverse flow field $\rho(\xi)$ (with $\alpha$ the angle between flow vector and parton propagation) have to be inferred from a dynamical evolution model. For ease of comparison with published results, we use a 2-d hydrodynamical evolution model \cite{Hydro} as in \cite{Correlations2} and a 3-d hydrodynamical model \cite{Hydro3d} as in \cite{Dyn01}.

In a static medium, $\hat{q}(\xi) = const.$ and thus Eq.~(\ref{E-omega}) exhibits the quadratic pathlength dependence for the energy scale parameter $\omega_c$ of radiative energy loss.

We do not attempt to construct a first-principles model of elastic energy loss in the following. Such an attempt faces a number of difficulties, among them the running of the strong coupling $\alpha_s$ into a soft regime, the treatment of interactions of the high-$p_T$ parton in the hadronic phase of the medium, the precise nature of scattering partners in the QGP phase and $q \leftrightarrow g$ conversion reactions. Instead, we try to construct a class of models which has the expected parametric dependence of energy loss probability distributions on the pathlength and infer then from the data limits for the parameters which characterize the model class. Since any observable considered here involves massive geometrical averaging (Eqs.~(\ref{E-P_TAA}),(\ref{E-P_phi})) and since furthermore the nuclear suppression factor is not very sensitive to the shape of even the averaged $\langle P(\Delta E) \rangle_{T_{AA}}$ \cite{gamma-hadron}, we may safely assume that details in the modelling of $P(\Delta E)_{path}$ will not influence the outcome significantly.

Note that in an ideal quark-gluon plasma (QGP), $\epsilon^{3/4}(\xi) (\cosh \rho(\xi) - \sinh \rho(\xi) \cos\alpha)$ is a measure of the entropy density and hence (up to a numerical constant) of the number density $\rho_M$ of scattering centers which can be probed by a parton on its trajectory $\xi$. If the cross section for elastic energy loss is $\sigma_{el}$ and the medium density $\rho_M$, the number of scatterings $dN$ on a path $d\xi$ is given by $dN = \sigma \rho_M d\xi$. Exponentiating this expression, we expect that the escape probability $P_0$ without undergoing energy loss for a parton is parametrically given by

\begin{equation}
\label{E-P0}
P_0 = \exp\left[- const. \cdot \sigma_{el} \int \tilde{\rho}_M(\xi)d\xi \right] = \exp[- \gamma \cdot \kappa]
\end{equation}

where we have assumed that $\sigma_{el}$ is approximately independent of $\xi$ and $\kappa$ is defined in analogy with Eq.~(\ref{E-omega}) as

\begin{equation}
\label{E-kappa}
\kappa = \int d\xi \epsilon^{3/4}(\xi) (\cosh \rho(\xi) - \sinh \rho(\xi) \cos\alpha)
\end{equation}

taking into account the flow corrections to the probed density. Here $\gamma$ is a parameter with dimensions of a cross section measuring the interaction strength, and hence $\gamma_g = 9/4 \gamma_q$ must hold to account for the different color factors of quarks and gluons.

Note that in the case of radiative energy loss as calculated using the assumption of \cite{Jet1,QuenchingWeights} the escape probability is determined by the condition of no radiation beyond the vacuum shower evolution rather than by a no-scattering condition. It can then be cast into the form \cite{QuenchingWeights}

\begin{equation}
P_0^{rad} =  \lim_{\nu \rightarrow \infty} \exp \left[ -\int_0^\infty d\omega \frac{dI(\omega)}{d\omega} \left( 1-e^{-\nu \omega}\right)\right]
\end{equation}

which requires the knowledge of the spectrum of medium-induced radiation $\omega \frac{dI(\omega)}{d\omega}$ as a function of radiated energy $\omega$. This in turn depends on a decoherence condition for radiated quanta as well as LPM interference and cannot easily
be cast into a simple form comparable to Eq.~(\ref{E-P0}). A good discussion of the underlying physics can be found in \cite{QuenchingWeights}.

If the parton does not escape without energy loss, it must undergo a shift in energy (there is also the possibility that a strong shift into a thermal regime occurs, which is equivalent to an absorption of the parton). It is reasonable to assume that the mean value of the shift in energy will grow linear in the number of scatterings $N$ as 
\begin{displaymath}
d\Delta E = \Delta E_{1} \sigma_{el} \rho_M d\xi
\end{displaymath}

with $\Delta E_1$ the mean energy loss per scattering whereas the fluctuations around the mean will grow like $\sqrt{N}$. Assuming a Gaussian distribution, this leads to the ansatz

\begin{equation}
\label{E-Elastic}
P(\Delta E)_{path} = P_0 \delta(\Delta E) + \mathcal{N} \exp\left[ \frac{(\Delta E - \alpha \kappa)^2}{\beta \kappa}  \right]
\end{equation}

where $\mathcal{N}$ is a normalization such that $\int_0^\infty P(\Delta E) = 1$ and (\ref{E-Elastic}) has to hold for quarks and gluons separately due to the different color factor. $\alpha$ is a parameter with the dimensions of a cross section times the energy shift per reaction. 

This class of energy loss models is characterized by three parameters:

\begin{itemize}
\item $\alpha$ controls the mean shift in energy per expected scattering
\item $\beta$ governs the strength of fluctuations around this mean shift. If $\beta$ is small, the model will have a strong correlation between path (and hence initial vertex) and shift in energy, if the parameter is large, this correlation is lessened
\item $\gamma$ finally determines the magnitude of the escape probability.
 \end{itemize}

In a microscopical model, the actual distribution would presumably not be strictly Gaussian and the three parameters would be correlated and calculable. Here, we will however pursue a different approach and see what can be inferred from the data.

\section{Determining the Parameter Range}

\begin{figure*}[htb]
\epsfig{file=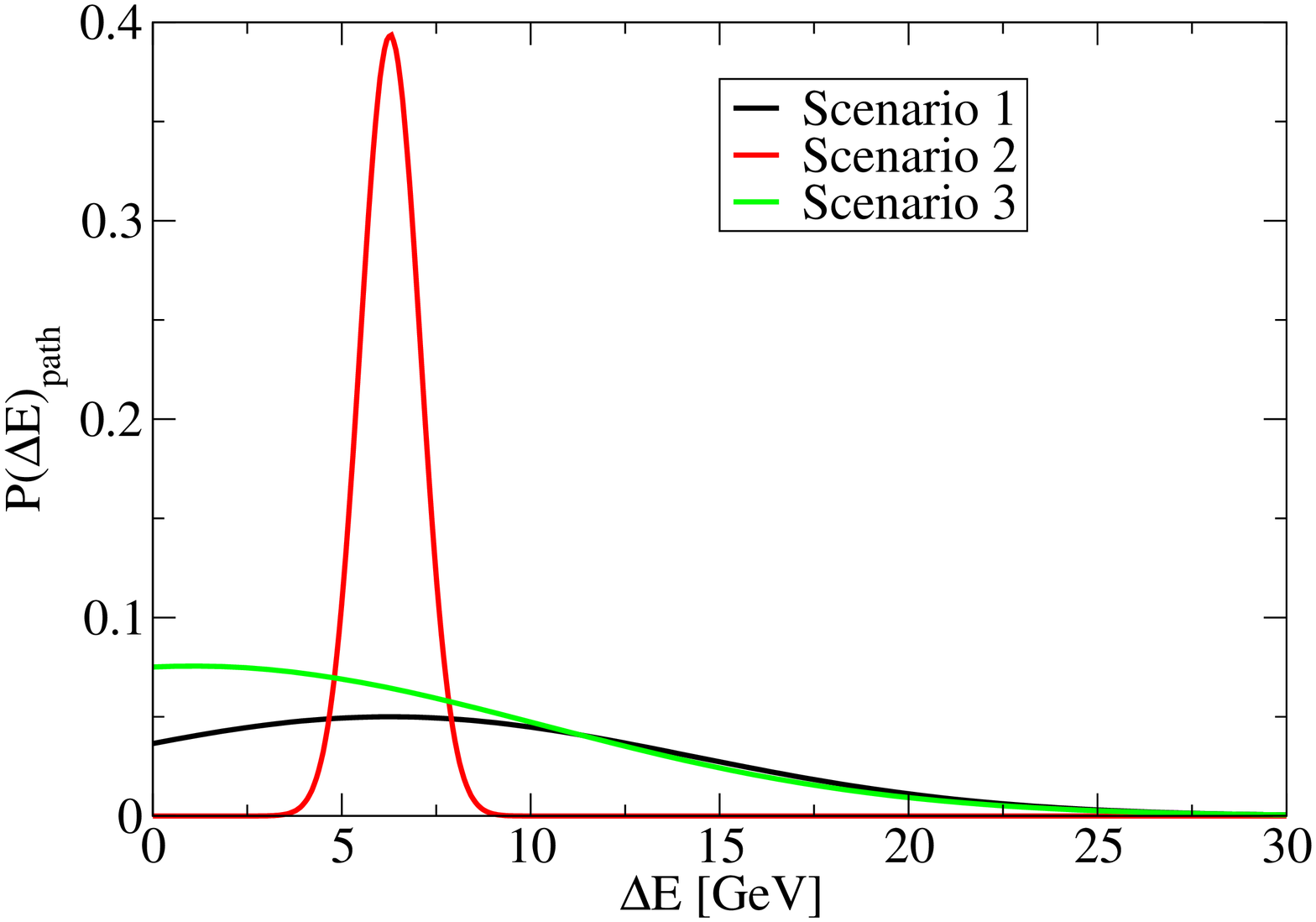, width=7cm} \epsfig{file=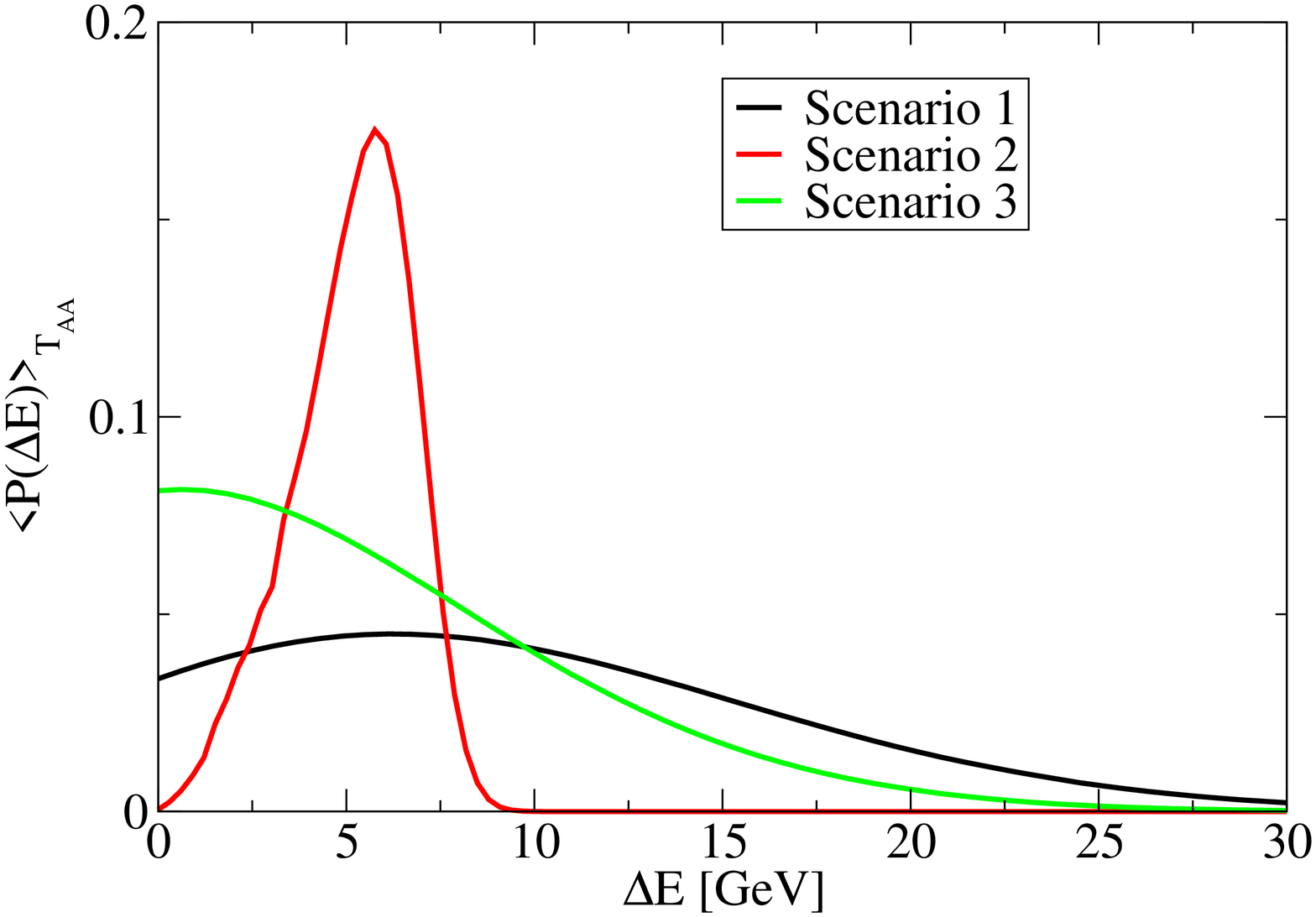, width=7cm}
\caption{\label{F-P_q}(Color online) Left panel: Continuous energy loss probability distribution for a quark propagating from the medium center outward in three different scenarios for elastic energy loss (see text). Right panel: Collision-geometry averaged energy loss distribution for quarks for three different scenarios.}
\end{figure*}

Our aim is to determine parameters of the elastic energy loss model such that $R_{AA}$ for central collisions is described well and to use these parameter settings to compute observables where a different geometrical avergaing is performed.
It does not seem straightforward to choose from the model space given by all possible $(\alpha, \beta, \gamma)$ all the solutions which are compatible with $R_{AA}$. However, according to \cite{gamma-hadron}, at least one parameter is fixed by the normalization of $R_{AA}$.

Let us consider $\gamma$ first: Clearly, the averaged discrete escape probability $\langle P_0 \rangle$ may not be larger than $R_{AA}$. On the other hand, if it is smaller than the error on $R_{AA}$, this term is irrelevant. Via 

\begin{equation}
\frac{1}{2\pi} \int_0^{2\pi}  
\negthickspace \negthickspace \negthickspace d\phi 
\int_{-\infty}^{\infty} \negthickspace \negthickspace \negthickspace \negthickspace dx_0 
\int_{-\infty}^{\infty} \negthickspace \negthickspace \negthickspace \negthickspace dy_0 P(x_0,y_0)  
\exp[-\gamma \kappa] < R_{AA}
\end{equation}

this translates to (evolution-model dependent) constraints on $\gamma$.

Let us now consider a scenario which is chosen such that $\langle P_0 \rangle$ is large and close to $R_{AA}$. It follows that there must be a constraint on the shape of the continuous shift probability $P(\Delta E)_{path}$. The essential idea is apparent from the following:

Averaging over both parton species and neglecting the fragmentation (which is a subleading correction on the shape of $R_{AA}$ in the high $p_T$ region \cite{ppbar}), $R_{AA}$ can be estimated as

\begin{equation}
\label{E-R_AA_approx}
R_{AA}(p_T) \approx \langle P_0 \rangle_{T_{AA}} + \int_{0+\epsilon}^{E_{max}} \negthickspace \negthickspace \negthickspace \langle P(\Delta E)_{T_{AA}} \rangle  \frac{\frac{dN_{part}}{dk_T}(p_T + \Delta E)}{\frac{dN_{part}}{dk_T}(p_T)} 
\end{equation}

where the lower integration boundary does not include the discrete contribution at 0 and $E_{max}$ is the kinematic limit for the parton energy. If the spectrum is approximated by a power law $\sim 1/k_T^n$, then 

\begin{displaymath}
\frac{\frac{dN_{part}}{dk_T}(p_T + \Delta E)}{\frac{dN_{part}}{dk_T}(p_T)}  \approx 1 / (1 + \frac{\Delta E}{p_T})^n.
\end{displaymath}

In other words, $R_{AA}$ can be written as a constant term plus a term which increases with $p_T$ and corresponds to the integral of the energy loss probability density, weighted by a steeply falling spectrum. From this term, sizeable contributions to $R_{AA}$ will come if $\langle P(\Delta E)_{T_{AA}} \rangle$ contains a lot of strength close to $\Delta E = 0$ (in this case the weight factor is close to unity) or if the spectrum is flat, i.e. $n$ is small (as at the LHC \cite{LHC}) or if $p_T \gg \Delta E$, i.e. at high $p_T$.  Thus, there is the generic expectation of a rise of $R_{AA}$ with $p_T$ \cite{JyvProc}.

If $\gamma$ is already close to the allowed limit, $\alpha$ and $\beta$ must thus arrange in such a way that the shift term is small in order not to violate the limit. This means that the continuous energy loss distribution cannot contain much strength close to $\Delta E = 0$.

There are two possible ways this could happen (note that the continuous part must integrate to approximately $1-R_{AA}$): First, the distribution could be very flat and extend to large $\Delta E$, thus the contribution in any given fixed interval in $\Delta E$ would be small. Or second, the distribution could be comparatively narrow but peaked at some large $\Delta E$.

In the case of radiative energy loss, the first scenario is realized, and for RHIC kinematics $\langle P(\Delta E)\rangle_{T_{AA}}$ extends to $O(100)$ GeV \cite{Correlations2}. However, this cannot be so in the elastic case. Assume that along some fixed parton path elastic and radiative processes lead to the same mean energy loss. Then, a comparison of Eqs.~(\ref{E-omega}) and (\ref{E-kappa}) shows that parametrically the radiative energy loss will be smaller for all shorther paths but larger for all longer path. Thus, the dynamical range of expected energy loss generated in the model between short pathlength contributions close to the surface and long paths traversing the whole medium is vastly greater for radiative energy loss, making it very difficult to obtain a flat distribution in elastic energy loss unless $\beta$ is set to (unnaturally) large values. However, when this is done, the probability distribution even for a single path is very wide, i.e. large fluctuations destroy the position-energy loss correlation and hence the tomographic information.

We investigate two different scenarios with a large discrete escape probability (chosen such that it is comparable with the discrete quenching weights in the radiative energy loss scenario \cite{QuenchingWeights}). The parameters of each scenario can be found in table \ref{T-Params}. In the first one (Scenario 1) we assume a large value of $\beta$, i.e. sizeable fluctuations of the energy loss given a path, in the second one (Scenario 2) we choose small $\beta$. The continuous energy loss probability distribution using the medium evolution provided by the 3-d hydrodynamics for quarks for a single path from the medium center and and the geometry-averaged distribution $\langle P(\Delta E) \rangle_{T_{AA}}$ for both scenarios are shown in Fig.~\ref{F-P_q}.

We also investigate a third scenario (Scenario 3) in which the discrete escape probability is adjusted to half the value of scenario 1 and 2. By the same argument seen above, the integral term in Eq.~(\ref{E-R_AA_approx}) must contribute more, hence the probability distribution needs strength close to $\Delta E = 0$, leading to some $p_T$ dependent growth of $R_{AA}$. This can only be achieved by choosing a comparatively large $\beta$. The resulting change in the shape of $R_{AA}$ due to the increased rise with $p_T$  as compared with the other scenarios disfavours even larger values of $\gamma$ (or a complete absence of escape without energy loss).

\begin{table}
\begin{tabular}{cccc}
\hline
&Scenario 1 & Scenario 2 & Scenario3 \\
\hline \hline
$\alpha$ [GeV$^{-1}$] & 0.35 &0.35 & 0.04\\
$\beta$ & 7.0 & 0.07 & 6.4\\
$\gamma$ [GeV$^{-2}$] & 0.085 & 0.085 & 0.12\\
\hline
\end{tabular}
\caption{\label{T-Params}Parameters for the three different elastic energy loss scenarios described in the text}
\end{table}

\section{$R_{AA}$ vs. reaction plane}

\begin{figure*}[htb]
\epsfig{file=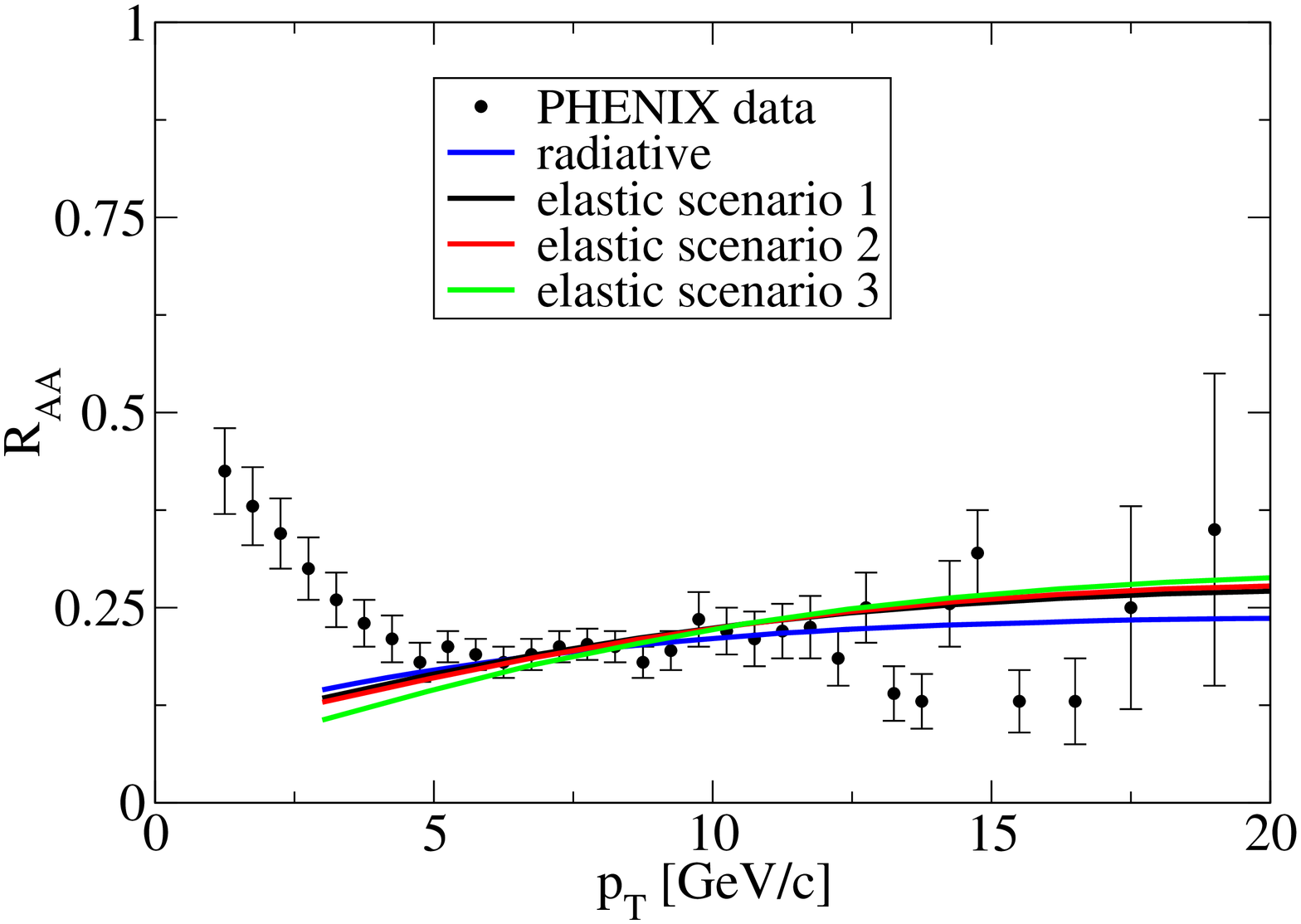, width=7cm} \epsfig{file=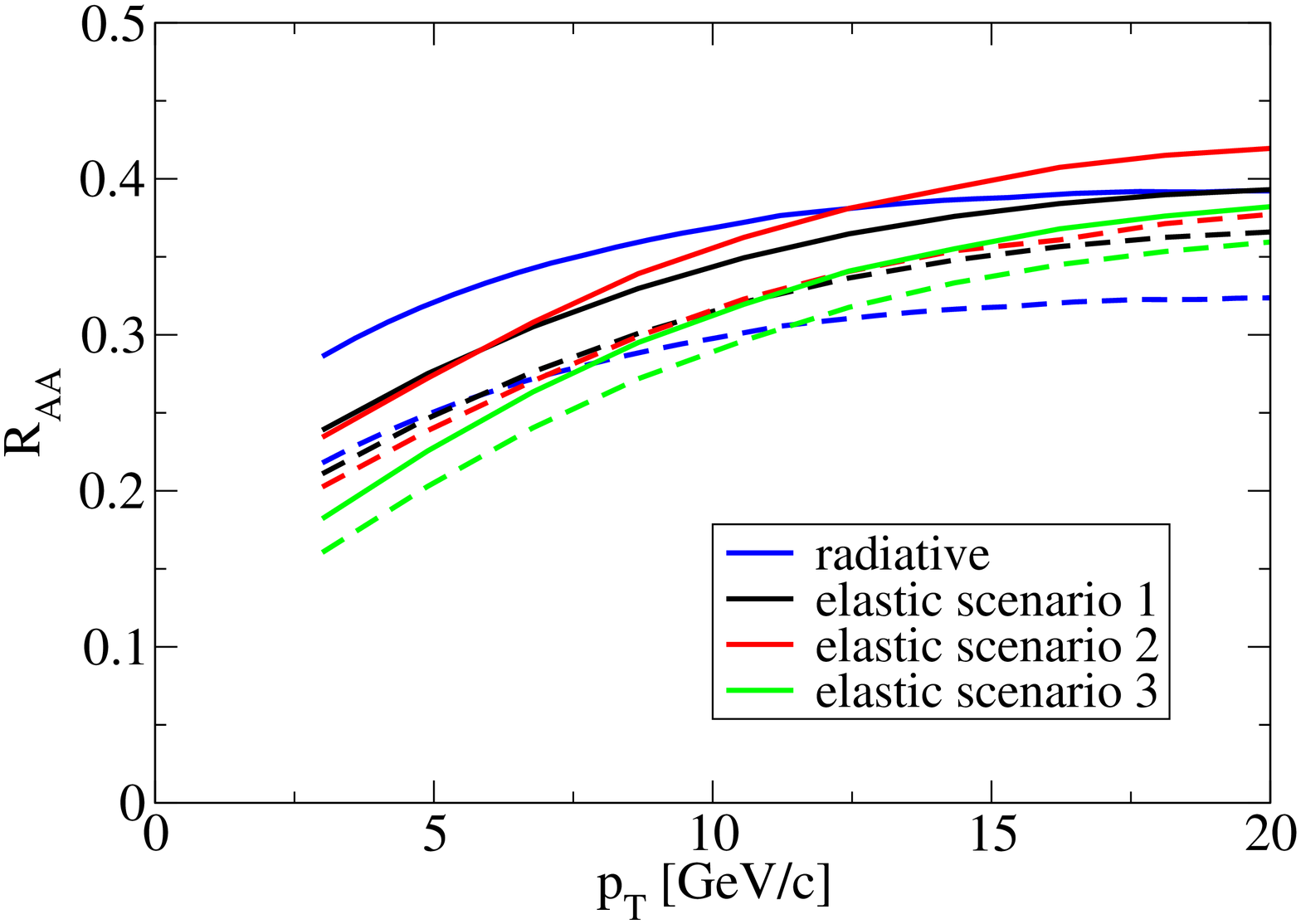, width=7cm}
\caption{\label{F-R_AA}(Color online) Left panel: Calculated nuclear suppression factor for radiative energy loss \cite{Dyn01} and for the three scenarios of elastic energy loss described in the text for central collisions (impact parameter $b=2.4$ fm) as compared to the PHENIX data \cite{PHENIX_R_AA} Right panel: As on left panel, except for $b=7.5$ fm. In-plane emission is indicated by solid lines, out-of-plane emission by dahsed lines. }
\end{figure*}

In Fig.~\ref{F-R_AA} we show the resulting $R_{AA}$ for different angles with respect to the reaction plane for central Au-Au collisions at 200 AGeV and for non-central collisions at impact parameter $b=7.5$ fm based on the medium description of the 3-d hydrodymnamics code \cite{Hydro3d}. For comparison we also include the radiative energy loss calculation.

By construction, differences between the scenarios are not very pronounced in central collisions. However, the extrapolation to non-central collisions is rather different in all cases. First, all three elastic energy loss scenarios exhibit a pronounced rise with $p_T$ which is stronger than in the radiative case. As outlined above, this is connected with the strength of $\langle P(\Delta E)\rangle _\phi$ close to $\Delta E =0$. However, second and possibly more impartant, the splitting between in-plane and out-of-plane emission in elastic energy loss is at most half of what is seen for radiative energy loss.

This is not an unecpected feature: In a dynamic medium, the leading initial density dilution goes as $1/\tau$ due to the longitudinal expansion (changing into $1/\tau^{c(T)}$ with $c(T)>1$ and growing later as pressure gradients drive accelerated expansion of the medium). Inserting this result int Eq.~(\ref{E-kappa}) shows that elastic energy loss is dominated by early times when the medium is dense. At late times, $\kappa$ increases only logarithmically. Using typical evolution models, this implies that energy loss models with a linear pathlength distribution lead to sizeable losses only in the first 2-3 fm/c evolution time and become insensitive to late time behaviour. However, if the medium becomes effectively transparent after 2-3 fm, a parton cannot probe density gradients much larger than this. This implies a loss of the sensitivity to in-plane vs. out-of-plane emission for some partons, and even more dramatic, a loss of sensitivity to the difference between average near-side and away side pathlength. Consistent with the expectation, the smallest splitting is observed when $\beta$ is set to large values and position-energy loss correlations are weakened in addition.

\section{Back-to-back dihadron correlations}

Using the MC code of \cite{Correlations2}, we compute the strength of hard hadronic back-to-back correlations for all three scenarios of elastic energy loss and compare with the STAR measurement \cite{Dijets1,Dijets2}.

\begin{figure}[htb]
\epsfig{file=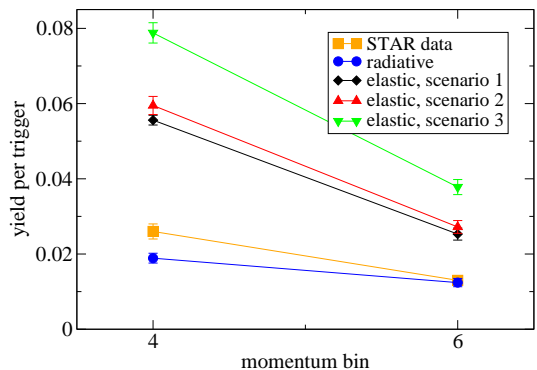, width=7cm}
\caption{\label{F-back-to-back}(Color online) Away side yield per trigger for central 200 AGeV Au-Au collisions for a trigger range between 8 and 15 GeV as a function of momentum bin as compared with STAR data \cite{Dijets1,Dijets2}. Shown is radiative energy loss in comparison with the three scenarios for elastic energy loss (see text) based on a 2-d hydrodynamical model for the medium evolution \cite{Hydro}.}
\end{figure}

As can easily be seen, elastic energy loss gives a parametrically wrong result --- the calculation overshoots the data by a factor of more than two. This expectation was already formulated in \cite{Correlations1}: In the radiative energy loss calculation, the difference between near side emission and away side emission arises from the drastic differences in average pathlength on near and away side \cite{Correlations2}. However, since in all elastic scenarios only pathlengths of 2-3 fm can be probed before the medium becomes effectively dilute, such mechanism cannot play a role. Thus, the main difference between near side and away side is that partons on the near side initially propagate into a zone of lower density and partons on the away side into higher density. This accounts for the fact that there is additional suppression on the away side, however it is by far not as strong as in the radiative case. If the position-energy loss correlation is further weakened (as in Scenario 3) the disagreement with the data is more pronounced.

\section{Discussion}

It seems clear that a scenario in which energy loss is exclusively elastic is not very realistic --- in reality, energy loss is presumably caused by different component processes which have different pathlength dependence. We may however estimate an upper bound for the strength of an elastic component with a simple ansatz

\begin{equation}
P(\Delta E)_{path} = f \cdot P(\Delta E)_{el} + (1-f) P(\Delta E)_{rad}
\end{equation}

where a fraction $f$ of the energy loss probability distribution is caused by an elastic channel.

The biggest uncertainty in the calculation of back-to-back correlations is given by the d-Au baseline for the correlation strength. If this uncertainty is combined with the uncertainty of the yield per trigger in Au-Au collisions and scenario 1 is chosen as being closest to the data, the 4-6 GeV momentum bin provides the strongest constraint and marginally allows an elastic energy loss fraction $f < 0.32$. However, this assumes that fragmentation is the only source of hadrons in this momentum bin which is not consistent with the fact that both radiative and collisional energy loss individually underpredict $R_{AA}$ in this momentum window (cf Fig.~\ref{F-R_AA}). Thus, the true upper limit on $f$ is even smaller. On the other hand, the most likely value of $f$ in the analysis comes out rather small as $f=0.086$.

In modelling the elastic energy loss, we have made several simplifying assumptions: First, we have assumed eikonal propagation of partons whereas in reality elastic energy loss leads to a deflection of partons. Furthermore, in writing down Eq.~(\ref{E-Elastic}) we have assumed that $\sigma_{el}$ doesn't strongly depend on the medium. However, if $\alpha_s(T)$ (and hence the interaction strength) grows very strongly for $T \rightarrow T_C$ this could to some extend effectively result in a deviations from a strictly linear pathlength dependence (the same effect would however occur for radiative energy loss). Furthermore, we have assumed a Gaussian distribution for the elastic energy loss probability. We have checked that the results do not crucially depend on this point. The reason is that the substantial geometrical averaging over many paths, cf. Eqs.~(\ref{E-P_TAA},\ref{E-P_phi}) erases any detailed information on the shape of the distribution given a single path. In particular, we verified that replacing a Gaussian shape by a box shape with the same r.m.s width does not alter the resulting $R_{AA}$ by more than 5\% and yields the same back-to-back correlations within the statistical errors of the MC simulation.

 Finally, we have neglected any energy dependence of the energy loss probability distributions and hence finite energy corrections. However, while these (and other) effects would clearly have an effect on the energy loss probabilities (to details of which the result is largely insensitive) it is less clear how they could possibly compensate for the  generically different pathlength dependence of the two regimes.

It appears that the general features of this analysis are rather robust and there is a physics reason why this should be so: The difference in near side and away side suppression is consistent with a quadratic pathlength dependence but not with a linear dependence of energy loss. As apparent from the rather consistent overprediction of the away side yield even for large discrete escape probability (the best possible scenario), this is a quite generic statement which holds for a whole class of elastic energy loss models and does not depend on details of how the probability distribution appears microscopically. It is hard to imagine a mechanism which would change this picture without strongly changing the parametric dependence on pathlength. This places rather stringent limits on the relative magnitude of elastic contributions to light quark and gluon energy loss which may help to constrain microscopical models of energy loss.

\begin{acknowledgments}
 
I'd like to thank Berndt M\"{u}ller and Kari Eskola for valuable discussions on the problem. This work was financially supported by the Academy of Finland, Project 115262. 
 
\end{acknowledgments}


\begin{thebibliography}{99}

\bibitem{Jet1}
  M.~Gyulassy and X.~N.~Wang,
  Nucl.\ Phys.\ B {\bf 420}, (1994) 583.
 
 
\bibitem{Jet2}
  R.~Baier, Y.~L.~Dokshitzer, A.~H.~Mueller, S.~Peigne and D.~Schiff,
  Nucl.\ Phys.\ B {\bf 484}, (1997) 265.
 
 
\bibitem{Jet3}
  B.~G.~Zakharov,
  JETP Lett.\  {\bf 65}, (1997) 615.
 
\bibitem{Jet4}
  U.~A.~Wiedemann,
  Nucl.\ Phys.\ B {\bf 588}, (2000) 303.
 
 
\bibitem{Jet5}
  M.~Gyulassy, P.~Levai and I.~Vitev,
  Nucl.\ Phys.\ B {\bf 594}, (2001) 371.
 
 
\bibitem{Jet6}
  X.~N.~Wang and X.~F.~Guo,
  Nucl.\ Phys.\ A {\bf 696}, (2001) 788.

\bibitem{Dainese}
  A.~Dainese, C.~Loizides and G.~Paic,
  Eur.\ Phys.\ J.\  C {\bf 38} (2005) 461.


\bibitem{Dyn01}
  T.~Renk, J.~Ruppert, C.~Nonaka and S.~A.~Bass,
  Phys.\ Rev.\  C {\bf 75} (2007) 031902.

\bibitem{Dyn02}
  A.~Majumder, C.~Nonaka and S.~A.~Bass,
  arXiv:nucl-th/0703019.

\bibitem{Dyn03}
  G.~Y.~P.~Qin, J.~Ruppert, S.~Turbide, C.~Gale, C.~Nonaka and S.~A.~Bass,
  arXiv:0705.2575 [hep-ph].

\bibitem{Correlations1}
  T.~Renk,
  Phys.\ Rev.\  C {\bf 74} (2006) 024903.

\bibitem{Correlations2}
  T.~Renk and K.~J.~Eskola,
  Phys.\ Rev.\  C {\bf 75} (2007) 054910.

\bibitem{Correlations3}
  H.~Zhang, J.~F.~Owens, E.~Wang and X.~N.~Wang,
  Phys.\ Rev.\ Lett.\  {\bf 98} (2007) 212301
  [arXiv:nucl-th/0701045].

\bibitem{ppbar}
  T.~Renk and K.~J.~Eskola,
  Phys.\ Rev.\  C {\bf 76} (2007) 027901.

\bibitem{HQPuzzle}
  M.~Djordjevic,
  J.\ Phys.\ G {\bf 32} (2006) S333
  [arXiv:nucl-th/0610054].

\bibitem{Mustafa}
  M.~G.~Mustafa,
  Phys.\ Rev.\  C {\bf 72} (2005) 014905.

\bibitem{Mustafa2}
  M.~G.~Mustafa and M.~H.~Thoma,
  Acta Phys.\ Hung.\  A {\bf 22} (2005) 93.

\bibitem{DuttMazumder}
  A.~K.~Dutt-Mazumder, J.~e.~Alam, P.~Roy and B.~Sinha,
  Phys.\ Rev.\  D {\bf 71} (2005) 094016.

\bibitem{Djordjevic}
  M.~Djordjevic,
  Phys.\ Rev.\  C {\bf 74} (2006) 064907.

\bibitem{Wicks}
  S.~Wicks, W.~Horowitz, M.~Djordjevic and M.~Gyulassy,
  Nucl.\ Phys.\  A {\bf 784} (2007) 426.

\bibitem{gamma-hadron}
  T.~Renk,
  Phys.\ Rev.\  C {\bf 74} (2006) 034906.

\bibitem{QuenchingWeights}
  C.~A.~Salgado and U.~A.~Wiedemann,
  Phys.\ Rev.\ D {\bf 68}, (2003) 014008.

\bibitem{KKP}
 B.~A.~Kniehl, G.~Kramer and B.~Potter,
  Nucl.\ Phys.\ B {\bf 582}, (2000) 514.

\bibitem{AKK}
  S.~Albino, B.~A.~Kniehl and G.~Kramer,
  Nucl.\ Phys.\ B {\bf 725} (2005) 181.

\bibitem{Flow1}
  R.~Baier, A.~H.~Mueller and D.~Schiff,
  nucl-th/0612068.

\bibitem{Flow2}
H.~Liu, K.~Rajagopal and U.~A.~Wiedemann,
  hep-ph/0612168.

\bibitem{Hydro}
  K.~J.~Eskola, H.~Honkanen, H.~Niemi, P.~V.~Ruuskanen and S.~S.~Rasanen,
  Phys.\ Rev.\ C {\bf 72} (2005) 044904.
 
\bibitem{Hydro3d}
  C.~Nonaka and S.~A.~Bass,
  Phys.\ Rev.\ C {\bf 75}, 014902 (2007).

\bibitem{LHC}
  T.~Renk and K.~J.~Eskola,
  arXiv:0705.1881 [hep-ph].

\bibitem{JyvProc}
  T.~Renk and K.~J.~Eskola,
  arXiv:0706.4380 [hep-ph].

\bibitem{PHENIX_R_AA}
  M.~Shimomura  [PHENIX Collaboration],
  nucl-ex/0510023.

\bibitem{Dijets1}
  D.~Magestro  [STAR Collaboration],
  nucl-ex/0510002; talk Quark Matter 2005.
 
\bibitem{Dijets2}
  J.~Adams {\it et al.}  [STAR Collaboration],
  nucl-ex/0604018.

\end{thebibliography}
\end{document}